\documentclass[conference,10pt]{IEEEtran}
\IEEEoverridecommandlockouts
\usepackage{cite}
\usepackage{amsmath,amssymb,amsfonts}
\usepackage{amsthm}
\usepackage{float}
\usepackage{dsfont}

\usepackage{mathtools}
\usepackage{mathabx}
\usepackage{upgreek}
\usepackage{bbm}

\usepackage[inline]{enumitem}

\usepackage{graphicx}
\usepackage{epstopdf} % Automatically converts .eps to .pdf during compilation
\usepackage{svg}

\usepackage{textcomp}
\usepackage{xcolor}
\usepackage{pgfplots}
\usepackage{microtype}      % microtypography
\usepackage[acronym, shortcuts]{glossaries}
\usepackage{ifthen}
\usepackage{xargs}
\usepackage{hhline}
\usepackage{bm}
\usepackage{breqn}
\usepackage{caption}
\usepackage{subcaption}
\usepackage{array} %better arrays and tabular
\usepackage{multirow} %single entry in multiple row elements in a tabular
\usepackage[para]{threeparttable} %add capability for additional notes to tables
\usepackage{booktabs} %nicer looking tabular 
\usepackage{balance} %equalize the columns on the last page with \balance command
\usepackage[inline]{enumitem}
\usepackage{cite}
\usepackage[framemethod=tikz]{mdframed}
\usepackage[vlined, ruled, linesnumbered, commentsnumbered]{algorithm2e}
\usepackage[acronym]{glossaries}
\usepackage{hhline}
\usepackage{bm}
\usepackage{ifthen}
\usepackage[utf8]{inputenc}
\usepackage{lipsum}
\usepackage{float}
\usepackage{comment}
\usepackage{tablefootnote}
\usepackage{array}
\usepackage{xfrac}
\usepackage{balance}
\usepackage{url}
\def\y{{\mathbf{y}}}
\def\a{{\mathbf{a}}}
\def\e{{\mathbf{e}}}
\def\A{{\mathbf{A}}}
\def\S{{\mathbf{S}}}
\def\I{{\mathbf{I}}}
\def\bm#1{{\boldsymbol{#1}}}
\def\gam{{\boldsymbol{\gamma}}}
\def\Sig{{\boldsymbol{\Sigma}}}
\def\hop{{\mathsf{H}}}
\def\R{{\mathbb{R}}}
\def\C{{\mathbb{C}}}
\def\D{{\mathbb{D}}}
\def\Sset{{\mathbb{S}}}
\DeclareMathOperator{\diag}{diag}
\DeclareMathOperator{\tr}{tr}
\DeclareMathOperator{\supp}{supp}

\begin{document}

\sloppy

\title{Covariance Learning for Direction-of-Arrival Estimation with Sparse Linear Arrays}

\author{\IEEEauthorblockN{Majdoddin Esfandiari\IEEEauthorrefmark{1},  Esa Ollila\IEEEauthorrefmark{1}, Geert Leus\IEEEauthorrefmark{2}, and Daniel P. Palomar\IEEEauthorrefmark{3}}
\IEEEauthorblockA{\IEEEauthorrefmark{1}Department of Information and Communications Engineering,
Aalto University, Espoo, Finland\\
\IEEEauthorrefmark{2}Department of Microelectronics, Delft University of Technology, Delft, The Netherlands\\
{\IEEEauthorrefmark{3}Department of Electronic and Computer Engineering, 
The Hong Kong University of Science and Technology, Hong Kong}\\
{Email: majdoddin.esfandiari@aalto.fi, esa.ollila@aalto.fi, g.j.t.leus@tudelft.nl, palomar@ust.hk}} }

\maketitle

% =========================================================================
\begin{abstract}
Covariance learning (CL) estimates source powers and sensor-noise variance by minimizing the Type‑II (marginal) negative log-likelihood of the array measurements. We develop both grid-based and gridless CL direction-of-arrival (DOA) estimators applicable to arbitrary sparse linear array (SLA) geometries.  
The grid-based DOA estimator minimizes the Type‑II objective on an angular grid via successive convex approximation (SCA) and estimates the noise variance by a scalar bisection search. The gridless DOA estimator then refines each grid-based DOA over the continuum using a Newton method. Simulations on a six-sensor nested array, including an underdetermined case with nine sources, show that the proposed DOA estimators outperform state-of-the-art methods, despite jointly estimating the sensor noise variance.
\end{abstract}


% =========================================================================
\vspace{-0.15cm}
\section{Introduction}
\label{sec:intro}

Direction-of-arrival (DOA) estimation is a core problem in array signal
processing, with applications in radar, sonar, and wireless
communications~\cite{vantrees2002,pesavento2023three}. Subspace-based
estimators such as MUSIC~\cite{schmidt1986}, root-MUSIC~\cite{barabell1983}
and ESPRIT~\cite{roy1989} offer high accuracy at modest computational cost. However,  with $M$ sensors in a uniform linear array (ULA), they can resolve at most $M - 1$ sources, and performance degrades significantly when sample sizes are small, the signal-to-noise ratio (SNR) is low, or sources are closely spaced or correlated~\cite{esfandiari2025generalized}. Sparse linear arrays (SLAs) circumvent the former.
Nested~\cite{pal2010nested}, coprime~\cite{vaidyanathan2011sparse} and minimum
redundancy arrays~(MRAs)~\cite{moffet1968} are designed so that the \emph{difference
co-array} of $M$ sensors contains $\mathcal{O}(M^{2})$ distinct lags, which
raises the number of identifiable sources accordingly.  A common approach to exploit these lags is to first vectorize the array covariance matrix into a single virtual snapshot, then restore its rank via spatial smoothing (SS) before applying a subspace estimator, as in co-array MUSIC~\cite{pal2010nested,wang2017coarray}.

A parallel line of research formulates DOA estimation as a sparse signal recovery
problem on an angular grid~\cite{malioutov2005}. Methods that model the sources
as random vectors and estimate their powers by minimizing a Type-II
(marginal) likelihood are particularly attractive, because that objective has
fewer and better behaved stationary points than its Type-I counterparts, in which the source vectors are estimated. Sparse
Bayesian learning (SBL)~\cite{tipping2001,wipf2007,gerstoft2016} and the
covariance fitting methods such as sparse iterative covariance-based estimation (SPICE)~\cite{stoica2011spice}, iterative adaptive approach 
(IAA)~\cite{yardibi2010} and sparse asymptotic minimum variance (SAMV)~\cite{abeida2013} belong to this category. Their common weakness is the bias introduced by
the finite grid, which dominates the error at high SNR.
Off-grid~\cite{yang2013ogsbi} and
gridless~\cite{yang2015gridless,yang2014spa,pote2023maximum} methods
address it, but the gridless formulations are, with few exceptions, tied to
the Vandermonde structure of the ULA, through polynomial
rooting or Toeplitz semidefinite programs.

In this paper, we take the Type-II route and develop both grid-based and gridless DOA estimators
for arbitrary SLA geometries within the covariance learning (CL) framework,
in which the source powers and the noise variance are estimated jointly by
successive convex approximation (SCA)~\cite{scutari2014,esfandiari2026adsca}. We refer to the proposed DOA estimators as grid-based CL-SCA and gridless CL-SCA. Numerical studies on a six-sensor nested array, including an
underdetermined scenario with nine sources, illustrate the superiority of the proposed methods
over the state-of-the-art DOA estimators, in particular the gridless CL-SCA.

\emph{Notation}: bold upper-case and lower-case letters denote matrices and
vectors. $(\cdot)^{\top}$, $(\cdot)^{*}$ and $(\cdot)^{\hop}$ denote
transpose, complex conjugate and Hermitian transpose, $\tr(\cdot)$ and
$|\cdot|$ the trace and the determinant, and $\Re\{\cdot\}$ and $\Im\{\cdot\}$
the real and imaginary parts. $[\,\cdot\,]_{+}=\max(\cdot,0)$, $\diag(\a)$ is
the diagonal matrix built from $\a$, and $[\cdot]_{ii'}$ is the $(i,i')$-th
entry of its argument.

% =========================================================================
\vspace{-0.3cm}
\section{System Model}
\label{sec:model}

Consider an SLA whose $M$ omnidirectional sensors are located at the
positions
$\Sset=\{d_{1},\dots,d_{M}\}\subset\{0,1,\dots,\bar{M}-1\}$, measured in
units of half a wavelength, with $d_{1}=0$ and $d_{M}=\bar{M}-1$. The array receives $K$ narrowband far-field
signals from the distinct directions $\theta_{k}$, each parameterized by its
spatial frequency $\omega_{k}=-\pi\sin(\theta_{k})\in(-\pi,\pi]$, so that the
array response reads
\begin{equation}
  \a(\omega)=\big[e^{\jmath d_{1}\omega},\,e^{\jmath d_{2}\omega},\,\dots,\,
                  e^{\jmath d_{M}\omega}\big]^{\top}\in\C^{M}.
  \label{eq:steering}
\end{equation}
The signal observed by the
array at the time instant $l$ is expressed as
\begin{equation}
  \y_{l}=\sum_{k=1}^{K}\a(\omega_{k})\,x_{kl}+\e_{l},
  \qquad l=1,\dots,L,
  \label{eq:snapshot}
\end{equation}
where $x_{kl}$ is the signal emitted by the $k$-th source and
$\e_{l}\in\C^{M}$ is a zero-mean circular complex Gaussian noise vector with covariance matrix
$\mathrm{cov}(\e_{l})=\sigma^{2}\I$, independent of the sources. Modeling
the source vectors also as independent zero-mean circular complex Gaussian vectors gives
$\y_{l}\sim\mathcal{CN}_{M}(\mathbf{0},\Sig)$ with
\begin{equation}
  \Sig=\sum_{k=1}^{K}\gamma_{k}\,\a(\omega_{k})\a(\omega_{k})^{\hop}
       +\sigma^{2}\I ,
  \label{eq:cov}
\end{equation}
where $\gamma_{k}$ is the power of the $k$-th source.

\subsection{Grid-based covariance learning}
\label{subsec:cl}

Discretizing the angular interval $[-90^{\circ},90^{\circ})$ into $N\gg K$
points $\bar{\theta}_{1},\dots,\bar{\theta}_{N}$ and writing
$\bar{\omega}_{n}=-\pi\sin(\bar{\theta}_{n})$, we form the overcomplete
dictionary $\A=[\a(\bar{\omega}_{1}),\ldots,\a(\bar{\omega}_{N})] = [\a_{1},\ldots,\a_{N}]
\in\C^{M\times N}$, so that \eqref{eq:cov} can be recast as
$\Sig=\A\diag(\gam)\A^{\hop}+\sigma^{2}\I$ with a $K$-sparse source power vector
$\gam\in\R_{\geq0}^{N}$, given that sources lie on the grid. Recovering $\supp(\gam)$ is then equivalent to
estimating the DOAs. The source powers and the noise variance are obtained by
minimizing the Type-II negative log-likelihood function of the data, that is,
\begin{equation}
  \begin{array}{ll}
    \underset{\gam,\,\sigma^{2}}{\text{minimize}} &
      \ell(\gam,\sigma^{2})=\tr\!\big(\Sig^{-1}\S\big)+\log|\Sig| \\[2pt]
    \text{subject to} & \gam\geq\mathbf{0},\quad \sigma^{2}>0,
  \end{array}
  \label{eq:typeII}
\end{equation}
where $\S=L^{-1}\sum_{l}\y_{l}\y_{l}^{\hop}$ is the sample covariance matrix
(SCM). The first term of the objective is convex in $(\gam,\sigma^{2})$ and
the second is concave, so \eqref{eq:typeII} is not convex.

\subsection{The co-array structure of the objective}
\label{subsec:coarray}

Let us introduce
\begin{equation}
  \D=\{\,d_{i}-d_{i'}\;:\;i,i'\in\{1,\dots,M\}\,\}
  \label{eq:coarray}
\end{equation}
to denote the difference co-array of $\Sset$, let $\D^{+}=\D\cap\mathbb{Z}_{>0}$,
and let $U$ be the largest integer such that $\{-U,\dots,U\}\subseteq\D$. For
a nested array with $M_{1}$ inner and $M_{2}$ outer sensors one has
$U=M_{2}(M_{1}+1)-1$, which grows as $\mathcal{O}(M^{2})$~\cite{pal2010nested}. A CL estimator accesses the full
co-array aperture implicitly,  and therefore
identifies up to $U$ sources~\cite{pal2010nested,wang2017coarray} while
operating throughout on the $M\times M$ matrices $\Sig$ and $\S$.

% =========================================================================
\vspace{-0.15cm}
\section{Proposed Method}
\label{sec:method}

\subsection{Grid-based CL-SCA: source powers and noise variance}
\label{subsec:gridbased}

Problem \eqref{eq:typeII} is not convex, but it is the sum of the convex term
$g(\gam,\sigma^{2})=\tr(\Sig^{-1}\S)$ and the concave term
$f(\gam,\sigma^{2})=\log|\Sig|$. We therefore minimize it by SCA~\cite{scutari2014}, linearizing only the concave term around the
current iterate and keeping the convex term exactly. At iteration $v+1$ this
assigns to the $n$-th power the surrogate
\begin{align}
  \tilde{\ell}_{n}(\gamma_{n}\mid\gam^{v})
  &= g\big(\gamma_{n},\gam^{v}_{\setminus n},(\sigma^{v})^{2}\big) \nonumber
  \\ &+\nabla_{\!\gamma_{n}}f\big(\gam^{v},(\sigma^{v})^{2}\big)\,
    \big(\gamma_{n}-\gamma_{n}^{v}\big),
  \label{eq:surrogate}
\end{align}
where $\gam^{v}_{\setminus n}$ collects the powers other than the $n$-th one.
Since $f$ is concave, \eqref{eq:surrogate} majorizes the objective along the
$n$-th coordinate, with equality at $\gamma_{n}=\gamma_{n}^{v}$. Writing
$\Sig^{v}=\A\diag(\gam^{v})\A^{\hop}+(\sigma^{v})^{2}\I$ and
$\mathbf{b}_{n}^{v}=(\Sig^{v})^{-1}\a_{n}$, the matrix inversion lemma gives
$\nabla_{\!\gamma_{n}}f=\a_{n}^{\hop}\mathbf{b}_{n}^{v}$ and turns
\eqref{eq:surrogate} into a scalar convex function of $\gamma_{n}$ whose
minimizer over $\gamma_{n}\geq0$ is available in closed form \cite{CLSCA-SPAWC26},
\begin{equation}
  \hat{\gamma}_{n}(\gam^{v})=
  \Bigg[\gamma_{n}^{v}
        +\sqrt{\frac{[\mathbf{b}_{n}^{v}]^{\hop}\S\,\mathbf{b}_{n}^{v}}
                    {(\a_{n}^{\hop}\mathbf{b}_{n}^{v})^{3}}}
        -\frac{1}{\a_{n}^{\hop}\mathbf{b}_{n}^{v}}\Bigg]_{+} ,
  \label{eq:sca_update}
\end{equation}
$n=1,\dots,N$.\footnote{The derivation is given in~\cite[Sec.~III]{esfandiari2026adsca}.}
All $N$ powers are therefore updated simultaneously, at a cost of
$\mathcal{O}(NM^{2})$ flops, and the iterates are smoothed as
$\gam^{v+1}=\gam^{v}+\eta^{v}\big(\hat{\gam}(\gam^{v})-\gam^{v}\big)$ with the
diminishing step size $\eta^{v}=\eta^{v-1}(1-\epsilon\,\eta^{v-1})$, which is
what guarantees convergence within the SCA framework~\cite{scutari2014}.
Nothing in \eqref{eq:sca_update} uses the geometry of the array beyond the fact
that the columns of $\A$ are array responses.

The noise variance is treated as one more block of the same SCA iteration.
Let $\tilde{\gam}^{v}$ denote a $K$-sparse vector extracted from $\gam^v$. Its nonzero indices are  the support of the $K$ largest peaks of $\gam^v$, and the value at each selected index equals the sum of the entries of $\gam^v$ that comprise the corresponding peak. Linearizing $f$ with respect to $\sigma^{2}$, using
$\nabla_{\!\sigma^{2}}f=\tr(\Sig^{-1})$, and replacing $\gam^{v}$ with $\tilde{\gam}^{v}$, yields the surrogate 
\begin{align}
  \tilde{\ell}_{N+1}\big(\sigma^{2}\mid\gam^{v},(\sigma^{v})^{2}\big)
  &=\tr\!\big((\tilde{\Sig}_{0}^{v}+\sigma^{2}\I)^{-1}\S\big) \nonumber
  \\ &+\sigma^{2}\tr\!\big((\tilde\Sig^{v})^{-1}\big),
  \label{eq:sigma_surrogate}
\end{align}
up to an additive constant, where
$\tilde{\Sig}_{0}^{v}=\A\diag(\tilde{\gam}^{v})\A^{\hop}$ and $\tilde\Sig^{v} = \tilde{\Sig}_{0}^{v} + (\sigma^{v})^{2} \I$. Setting the
derivative of \eqref{eq:sigma_surrogate} to zero and using the eigenvalue decomposition
$\tilde{\Sig}_{0}^{v}=\mathbf{U}^{v}\bm{\Lambda}^{v}(\mathbf{U}^{v})^{\hop}$
reduces the update to the scalar equation
\begin{equation}
  \sum_{i=1}^{M}\frac{[\tilde{\S}^{v}]_{ii}}{(\lambda_{i}^{v}+\sigma^{2})^{2}}
  =\tr\big((\tilde\Sig^{v})^{-1}\big),
  \label{eq:sigma_eq}
\end{equation}
with $\lambda_{i}^{v}=[\bm{\Lambda}^{v}]_{ii}$ and
$\tilde{\S}^{v}=(\mathbf{U}^{v})^{\hop}\S\,\mathbf{U}^{v}$. Its left-hand side is
continuous and strictly decreasing on $\R_{>0}$, so \eqref{eq:sigma_eq} has at
most one positive root, which is retrieved by bisection on
$[\sigma^{2}_{\rm L},\tr(\S)/M]$ and then smoothed with the same step size
$\eta^{v}$.

\subsection{Gridless CL-SCA}
\label{subsec:gridless}

The grid-based CL-SCA returns $K$ frequencies $\omega_{1},\dots,\omega_{K}$
constrained to the grid, and results in a bias of the order of the grid
spacing. We remove this bias by refining each estimate on the continuum. Let
$\bar{\Sig}=\sum_{i=1}^{K}\gamma_{i}\a(\omega_{i})\a(\omega_{i})^{\hop}+\sigma^{2}\I$
and $\bar{\Sig}_{\setminus d}=\bar{\Sig}-\gamma_{d}\a(\omega_{d})
\a(\omega_{d})^{\hop}$, and isolate the contribution of the $d$-th source to the SCA surrogate. The optimization problem for refining the $d$-th source power and frequency can then be written, following \eqref{eq:surrogate}, as
\begin{equation}
(\gamma_d^{\rm opt},\omega_d^{\rm opt}) = \arg\min_{\omega  \in (-\pi,\pi]} \, \min_{\gamma  \geq 0} \, \bar{\ell}_d\left(\gamma,\omega  \right)   \label{eq:Gridless_1}
\end{equation}
where $\bar{\ell}_d\left(\gamma,\omega  \right) = \gamma \, p(\omega) - \left(\gamma \, s(\omega)\right)/\left(1+\gamma \, v(\omega) \right) $, and
\begin{equation}
  \begin{aligned}
    p(\omega)&=\a(\omega)^{\hop}\bar{\Sig}^{-1}\a(\omega), \\
    s(\omega)&=\a(\omega)^{\hop}\bar{\Sig}_{\setminus d}^{-1}\,\S\,
               \bar{\Sig}_{\setminus d}^{-1}\a(\omega), \\
    v(\omega)&=\a(\omega)^{\hop}\bar{\Sig}_{\setminus d}^{-1}\a(\omega).
  \end{aligned}
  \label{eq:pqv}
\end{equation}
Here, rather than joint optimization, we minimize $\bar{\ell}_d\left(\gamma,\omega  \right)$ over $\gamma$ analytically first. For fixed $\omega$, $\bar{\ell}_d\left(\gamma,\omega  \right)$ in \eqref{eq:Gridless_1} is convex in $\gamma$, and therefore any stationary point is the global minimum. Setting the first derivative of $\bar{\ell}_d\left(\gamma,\omega  \right)$ w.r.t. $\gamma$ to zero, we obtain $\gamma_{d}^{\mathrm{opt}}=[(\sqrt{s(\omega)/p(\omega)}-1)/v(\omega)]_{+}$. Substituting $\gamma_d^{\rm opt}$ for $\gamma$ in \eqref{eq:Gridless_1}, we obtain
\vspace{-0.1cm}
\begin{equation}
  \omega_{d}^{\mathrm{opt}}
  =\arg\max_{\omega\in(-\pi,\pi]}\;
   J_{d}(\omega):=\frac{\left(\sqrt{s(\omega)}-\sqrt{p(\omega)}\right)^{2}}
                       {v(\omega)}.
  \label{eq:Jd}
\end{equation}
A solution of \eqref{eq:Jd} based on a Newton method with backtracking is:
\begin{equation}
  \omega_{d}^{[q]}=\omega_{d}^{[q-1]}
   -\beta\,\frac{J_{d}'(\omega_{d}^{[q-1]})}{J_{d}''(\omega_{d}^{[q-1]})},
  \label{eq:newton}
\end{equation}
where $\beta$ is halved until $J_{d}(\omega_{d}^{[q]})\geq
J_{d}(\omega_{d}^{[q-1]})$, which prevents both overshoot and steps of
incorrect curvature. The derivatives $J_{d}'$ and $J_{d}''$ follow from those
of $p$, $s$ and $v$ by the chain rule. 

The efficiency of \eqref{eq:newton} rests on how the quadratic forms
\eqref{eq:pqv} are evaluated. For a general Hermitian matrix $\mathbf{X}$,
\begin{equation}
  \a(\omega)^{\hop}\mathbf{X}\,\a(\omega)
  =\sum_{i=1}^{M}\sum_{i'=1}^{M}[\mathbf{X}]_{ii'}\,
    e^{\jmath(d_{i'}-d_{i})\omega},
  \label{eq:qform_raw}
\end{equation}
so that, grouping the terms according to the value of the lag
$m=d_{i'}-d_{i}$ and defining the \emph{co-array lag sums}
\begin{equation}
  \tau_{\mathbf{X}}(m)=\!\!\sum_{(i,i'):\,d_{i'}-d_{i}=m}\!\![\mathbf{X}]_{ii'},
  \qquad m\in\D,
  \label{eq:lagsum}
\end{equation}
the quadratic form and its first two derivatives become the trigonometric
sums
\begin{equation}
  \begin{aligned}
    \a(\omega)^{\hop}\mathbf{X}\a(\omega)
      &=\tau_{\mathbf{X}}(0)
        +2\!\!\sum_{m\in\D^{+}}\!\!\Re\{\tau_{\mathbf{X}}(m)e^{\jmath m\omega}\},\\
    \tfrac{\partial}{\partial\omega}\,\a(\omega)^{\hop}\mathbf{X}\a(\omega)
      &=-2\!\!\sum_{m\in\D^{+}}\!\! m\,\Im\{\tau_{\mathbf{X}}(m)e^{\jmath m\omega}\},\\
    \tfrac{\partial^{2}}{\partial\omega^{2}}\,\a(\omega)^{\hop}\mathbf{X}\a(\omega)
      &=-2\!\!\sum_{m\in\D^{+}}\!\! m^{2}\,\Re\{\tau_{\mathbf{X}}(m)e^{\jmath m\omega}\}.
  \end{aligned}
  \label{eq:derivs}
\end{equation}
Applying \eqref{eq:lagsum}--\eqref{eq:derivs} to
$\mathbf{X}\in\{\bar{\Sig}^{-1},\,\bar{\Sig}_{\setminus d}^{-1}\S
\bar{\Sig}_{\setminus d}^{-1},\,\bar{\Sig}_{\setminus d}^{-1}\}$ yields
$p,s,v$ and their derivatives, and hence $J_{d}$, $J_{d}'$ and $J_{d}''$, in
$\mathcal{O}(|\D^{+}|)$ flops.  The lag sums can be formed at a cost of
$\mathcal{O}(M^{2})$ per source and refinement sweep.  
The noise
variance is re-estimated after every refinement sweep by finding the root of 
\eqref{eq:sigma_eq}, which costs
$\mathcal{O}(M^{2}K)$ flops by the low-rank form of
$\A\diag({\gam})\A^{\hop}$. The overall procedure of the proposed grid-based and gridless CL-SCA estimators is summarized in Algorithm~\ref{alg:clsca}. It
costs $\mathcal{O}(NM^{2})$ flops per grid-based iteration and
$\mathcal{O}(K(M^{3}+Q|\D^{+}|))$ flops per refinement sweep, so its complexity
is dominated by the grid-based stage whenever $N\gg K$.

\begingroup
\LinesNotNumbered

\begin{algorithm}[t]
\DontPrintSemicolon
\SetKwInOut{Input}{Input}
\SetKwInOut{Output}{Output}
\SetInd{0.4em}{0.4em}

% Caption formatting
\SetAlgoCaptionSeparator{\ }
\SetAlCapSty{textbf}
\SetAlCapNameSty{textnormal}

\Input{$\S$, $\A$, $\bar{\bm{\theta}}$, $\Sset$, $K$,
       $\{\eta^{v}\}$, $I_{\max}$, ${\rm ITER}$, $Q$, $\beta$.}

\nl
$\gam^{0}=\mathbf{0}$,
$(\sigma^{0})^{2}=\tr(\S)/(2M)$,
$\sigma^{2}_{\rm L}=10^{-8}$,
$\delta=10^{-4}$.\;

\BlankLine
\BlankLine
\textbf{Grid-based CL-SCA:}\;

\For{$v=0,\dots,I_{\max}-1$}{
  \nl
  $\Sig^{v}=\A\diag(\gam^{v})\A^{\hop}+(\sigma^{v})^{2}\I$,
  $\mathbf{B}^{v}=(\Sig^{v})^{-1}\A$,
  $\hat{\gamma}_{n}(\gam^{v})$ from \eqref{eq:sca_update},
  for $n=1,\dots,N$ in parallel,
  $\tilde{\gam}^{v}\leftarrow$ integrated power of the
  $K$ largest peaks of $\gam^{v}$,
  $(\sigma^{v+1})^{2}\leftarrow$ bisection on
  \eqref{eq:sigma_eq} smoothed by $\eta^{v}$,
  $\gam^{v+1}=\gam^{v}
  +\eta^{v}\big(\hat{\gam}(\gam^{v})-\gam^{v}\big)$.\;

  \lIf{$\|\gam^{v+1}-\gam^{v}\|_{2}<\delta$}{
    \textbf{break}
  }
}

\nl
$\bm{\theta}_{\rm G}\leftarrow$ $K$ DOAs corresponding
to the $K$ largest peaks of $\gam^{v+1}$,
$\gam_{\rm G}$,
$\boldsymbol{\omega}_{\rm G}
=-\pi\sin(\bm{\theta}_{\rm G})$.\;

\BlankLine
\BlankLine
\textbf{Gridless CL-SCA:}\;

\For{${\rm iter}=1,\dots,{\rm ITER}$}{
  \nl
  \For{$d=1,\dots,K$}{
    \nl
    form $\bar{\Sig}$, $\bar{\Sig}_{\setminus d}$
    and the lag sums \eqref{eq:lagsum}
    of the three matrices\;

    $Q$ Newton steps \eqref{eq:newton} on $J_{d}$,
    $\beta\leftarrow0.5\beta$
    while $J_{d}$ decreases\;

    $\gamma_{d}\leftarrow[(\sqrt{s/p}-1)/v]_{+}$
    at the new $\omega_{d}$\;
  }

  \nl
  Update $\sigma^{2}$ by finding the root of
  \eqref{eq:sigma_eq}.\;
}

\Output{$\bm{\theta}_{\rm G}$ and
        $\bm{\theta}_{\rm GL}=\arcsin(-\bm{\omega}/\pi)$.}

\caption{Grid-based and gridless CL-SCA DOA estimation for an SLA}
\label{alg:clsca}
\end{algorithm}

\endgroup

\vspace{-0.3cm}

%

%=========================================================================
%\vspace{-0.1cm}
\section{Simulation Results}
\label{sec:sim}

We consider a two-level nested array with $M_{1}=M_{2}=3$, i.e., 
$M=6$ physical sensors located at
$\Sset=\{0,1,2,3,7,11\}$ half-wavelengths, whose difference co-array is the
contiguous set $\D=\{-11,\dots,11\}$, so that $U=11$ sources are identifiable.
The proposed estimators are compared with seven state-of-the-art DOA
estimators that apply to sparse arrays: co-array spatial-smoothing MUSIC
(SS-MUSIC)~\cite{pal2010nested}, the iterative adaptive approach
(IAA)~\cite{yardibi2010}, SPICE~\cite{stoica2011spice},
SAMV~\cite{abeida2013}, off-grid sparse Bayesian inference (OGSBI)~\cite{yang2013ogsbi}, and the two
recent methods, namely SBL with grid refinement
(SBL-GR)~\cite{pote2023maximum} and the light-weight sequential SBL algorithm
(LWS-SBL)~\cite{pote2025lws}. 
The
stochastic Cram\'{e}r--Rao bound (CRB)~\cite{stoica1990} is shown as a
benchmark. All the on-grid methods, including the grid-based CL-SCA, use the same
uniform angular grid of $N=1801$ points over $[-90^{\circ},90^{\circ}]$.
OGSBI is an off-grid method that linearizes the manifold inside each cell and
estimates the residual offset, and it is therefore run on a coarse
$2^{\circ}$ grid, which is the regime it is designed for in~\cite{yang2013ogsbi}
and which we verified to be its most favorable setting here.
SBL-GR is given the same $2^{\circ}$ grid, since grid refinement exists
precisely to allow a coarse one.  
The SAMV power update is
damped by a factor of $0.3$, without which the parallel fixed-point iteration
diverges on a fraction of the realizations. All the other baselines are run
with the settings of the corresponding references. The noise variance is
treated as unknown by the proposed estimators and is estimated jointly with the
DOAs. For the proposed CL-SCA estimators, the step-size parameters are $\eta^{0}=0.99$ and $\epsilon=0.05$, the maximum
number of grid-based iterations is $500$ with the stopping rule
$\|\gam^{v+1}-\gam^{v}\|_{2}<10^{-4}$, the bisection in \eqref{eq:sigma_eq} is
run over $[10^{-8},\tr(\S)/M]$, and the refinement uses ten sweeps of at most
$Q=10$ Newton steps each with $\beta$ initialized to one and halved at each
backtracking step. 

The noise variance is  $\sigma^{2}=1$, the
first two sources have power $\sigma^{2}\cdot10^{\mathrm{SNR}/10}$ and the
remaining sources are $3$~dB stronger. All the results are averaged over
$1000$ Monte Carlo trials, and the performance metric is the root mean square
error (RMSE) in degrees, displayed on a logarithmic scale.

\begin{figure}[t]
\centering
\includegraphics[width=0.98\linewidth]{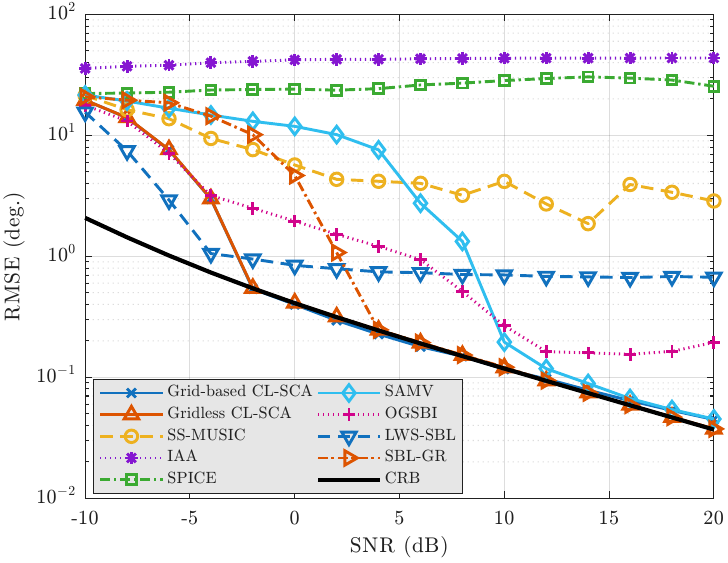}
\includegraphics[width=0.98\linewidth]{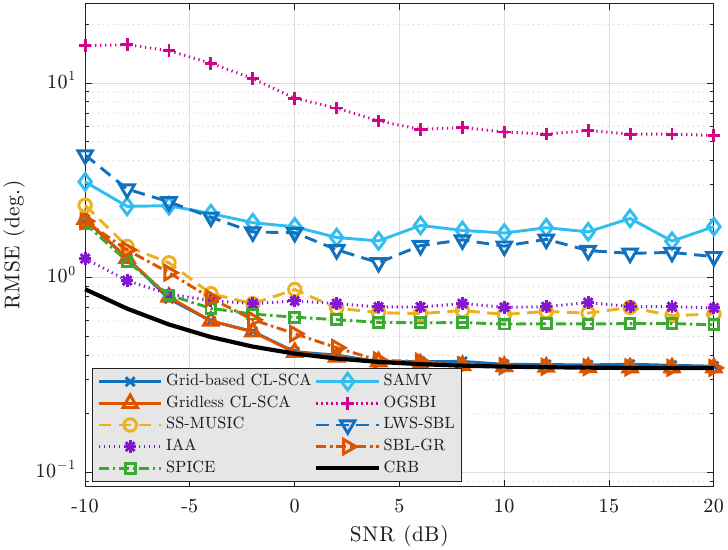}
\caption{RMSE versus SNR for uncorrelated sources
received by the $M=6$ nested array. (upper panel) $K=3$ sources with $\boldsymbol{\theta} = [-40.32^{\circ}, -34.68^{\circ}, 32.12^{\circ}] $ and $L=100$;
(lower panel) $K=9>M$ sources and $L=200$.}
\label{fig:snr}
\end{figure}

Fig.~\ref{fig:snr} shows the RMSE of the methods tested versus SNR. The upper panel considers $K=3$ uncorrelated sources with $\boldsymbol{\theta} = [-40.32^{\circ}, -34.68^{\circ}, 32.12^{\circ}] $ and
$L=100$, and the lower panel $K=9>M$ uncorrelated sources with $\boldsymbol{\theta} = [-55.82^{\circ}, -41.82^{\circ}, -27.82^{\circ}, -13.82^{\circ}, 0.18^{\circ}, 14.18^{\circ}, 28.18^{\circ}, \allowbreak 42.18^{\circ}, 56.18^{\circ}] $, i.e.,  the underdetermined regime. 
The lower panel uses $L=200$ and the CRB for the underdetermined case is shown as a benchmark \cite{wang2017coarray}. It can be observed that both the grid-based and gridless CL-SCA outperform the other methods tested in both scenarios considered in Fig.~\ref{fig:snr}.

\begin{figure}[t]
\centering
\includegraphics[width=0.98\linewidth]{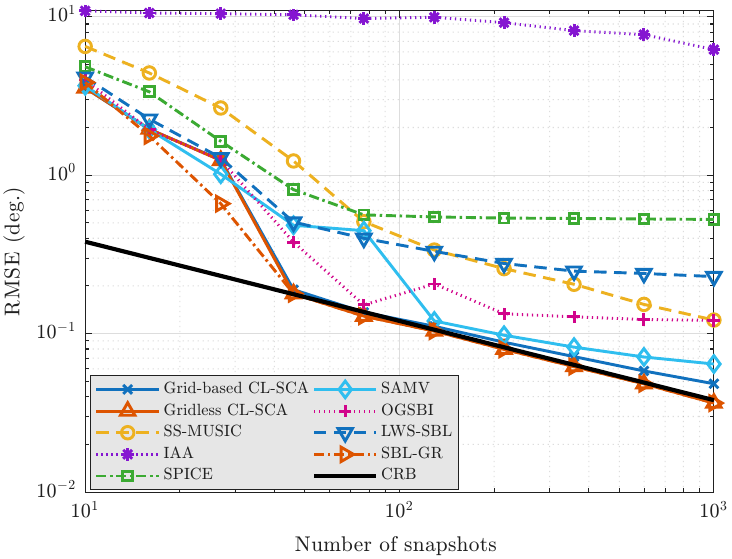}
\includegraphics[width=0.98\linewidth]{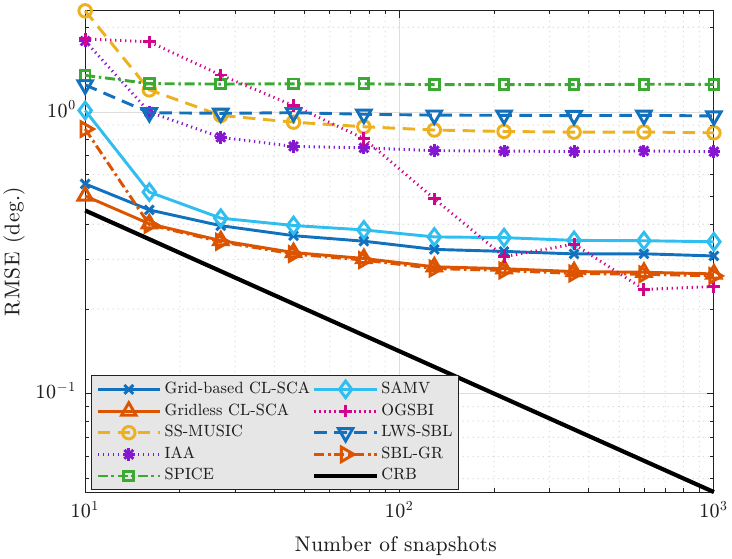}
\caption{RMSE versus the number of snapshots $L$ for $K=3$  sources $\boldsymbol{\theta} = [-28.32^{\circ}, -20.32^{\circ}, 10.12^{\circ}] $ received by the $M=6$ nested array at
$\mathrm{SNR}=5$~dB. (upper panel) uncorrelated sources; (lower panel) correlation coefficient
$\rho=0.95$ between the first two sources.}
\label{fig:snap}
\end{figure}

Fig.~\ref{fig:snap} shows the RMSE versus the number of snapshots $L$ at
$\mathrm{SNR}=5$~dB for $K=3$  sources $\boldsymbol{\theta} = [-28.32^{\circ}, -20.32^{\circ}, 10.12^{\circ}] $, for
uncorrelated sources and for a correlation coefficient of $\rho=0.95$ between
the first two sources. In the upper panel of Fig.~\ref{fig:snap}, gridless CL-SCA and SBL-GR deliver the best performance among the methods evaluated. Moreover, the bias of grid-based CL-SCA at large numbers of snapshots is eliminated by gridless CL-SCA. In the lower panel, gridless CL-SCA achieves the best performance. All methods, however, struggle in the presence of highly correlated sources.  

\vspace{-0.3cm}
\section{Conclusion}
\label{sec:concl}

We developed both grid-based and gridless CL-SCA  DOA estimators for sparse linear arrays.  
The latter is developed by refining the grid-based CL-SCA DOA estimates using a Newton method. 
Numerical studies on a six-sensor nested array show that the proposed grid-based and gridless CL-SCA estimators outperform state-of-the-art DOA methods, operate reliably with more sources than sensors, and that gridless CL-SCA removes the grid-mismatch
bias of grid-based CL-SCA at high SNR and/or large numbers of snapshots.

\section{Acknowledgments}
This work was supported by the Research Council of Finland under grant 368630 and the Hong Kong GRF 16206123 research grant. Beyond this support, the authors have no relevant financial or non-financial interests to disclose. Claude (Anthropic) was used to assist in writing the manuscript and coding the simulations. The authors take full responsibility for the results of this paper.

\section{Compliance with Ethical Standards}
This is a numerical simulation study for which no ethical approval was required.

\bibliographystyle{IEEEbib}

\end{document}